# Machine Learning Approach on Multiclass Classification of Internet Firewall Log Files

Md Habibur Rahman
Department of Computer Science and Engineering
*Daffodil International University*
Dhaka, Bangladesh
habibur15-12391@diu.edu.bd

Taminul Islam*
Department of Computer Science and Engineering
*Daffodil International University*
Dhaka, Bangladesh
taminul@ieee.org

Md Masum Rana
Department of Computer Science
*University of South Dakota*
Vermillion, SD, United States
mdmasum.rana@coyotes.usd.edu

Rehnuma Tasnim
Department of Computer Science and Engineering
*Daffodil International University*
Dhaka, Bangladesh
rehnuma15-12770@diu.edu.bd

Tanzina Rahman Mona
Department of Computer Science and Engineering
*Daffodil International University*
Dhaka, Bangladesh
tanzina15-2812@diu.edu.bd

Md. Mamun Sakib
Department of Computer Science and Engineering
*Daffodil International University*
Dhaka, Bangladesh
mamun15-12651@diu.edu.bd

*Abstract*— Firewalls are critical components in securing communication networks by screening all incoming (and occasionally exiting) data packets. Filtering is carried out by comparing incoming data packets to a set of rules designed to prevent malicious code from entering the network. To regulate the flow of data packets entering and leaving a network, an Internet firewall keeps a track of all activity. While the primary function of log files is to aid in troubleshooting and diagnostics, the information they contain is also very relevant to system audits and forensics. Firewall's primary function is to prevent malicious data packets from being sent. In order to better defend against cyberattacks and understand when and how malicious actions are influencing the internet, it is necessary to examine log files. As a result, the firewall decides whether to 'allow,' 'deny,' 'drop,' or 'reset-both' the incoming and outgoing packets. In this research, we apply various categorization algorithms to make sense of data logged by a firewall device. Harmonic mean F1 score, recall, and sensitivity measurement data with a 99% accuracy score in the random forest technique are used to compare the classifier's performance. To be sure, the proposed characteristics did significantly contribute to enhancing the firewall classification rate, as seen by the high accuracy rates generated by the other methods.

*Keywords— multiclass classification, internet firewall, log file, machine learning, networking*

## I. INTRODUCTION

By exchanging information online, your data might be subject to a variety of cyberattacks and breaches. As we enter a new era of information technology and the internet, the number of apps that may access data resources is growing at a dizzying rate. The log file contains entries that document system and application activity, as well as network activity for any connected devices. Log data is produced in vast quantities by all of the system's software and hardware components. In response to each conceivable occurrence on the internet, data is recorded in a log file [1]. This information was initially recorded for assistance in troubleshooting and diagnostics.

Firewalls are like toll booths for data packets on a network. System administrators install firewalls that are tailored to the needs of their specific business [2]. Firewalls have proven to be an integral component of modern communication networks due to the vital function they play in protecting the network from both external and internal threats [3]. Firewalls, in their most basic form, organize network log data in accordance with their rules, which may be defined manually or by default depending on particular criteria, such as the purpose of the link, which ports are effective communication and interpersonal, which subdivisions are permitted, etc. The organization that is using the firewall will determine its specific regulations [4]. In addition, keeping these rules up-to-date is a time-consuming and ongoing task, what with technological developments and the ever-evolving behavior of the environment. Actions, such as "Accept," "Drop," "Deny," or "Reset-both," are taken depending on these rules and many other aspects of the network log entries. Incorrectly handling a session might compromise security, leading to consequences like the loss of data or the inadvertent destruction of equipment, which could have a ripple effect on revenue.

The original purpose of log files was to document system activity. In the case of an attack or other malicious action, this can be used for forensics and audit trails [5]. Firewalls in a network determine whether or not traffic is allowed depending on policy by analyzing the data generated. For communications systems to operate smoothly and securely, firewall setup is essential. The efficient operation of a business's communication tools and other networked resources depends in large part on the setup of these systems. Firewalls are the electronic equivalent of security gates, restricting access to and from computer networks. The firewalls are set up by the system administrator to protect the company [6].

If hackers are able to break into a network's architecture, they may transfer data to unintended recipients or alter the data's veracity and consistency at any point in its existence. In response, several security measures, such as Internet Firewalls, Intrusion Detection/Prevention Systems (IDS/IPS) [7], and others, have been implemented at varying levels of protection to deal with security concerns.

The rest of our paper, we will review some related works on firewall log file system briefly in section II, then in section III will explain about the dataset, methods used to classify the dataset and comparison among the outputs, finally section IV concludes with the final result and future scopes in this field.

## II. LITERATURE REVIEW

This section will detail and analyze prior studies done in the field of internet gate log file categorization to provide additional context for the efficacy of the algorithm applied to the dataset to detect the discrepancy.

A machine learning automation technique was employed by Shridhar Allagi et al. [8]. In his study, he describes how he used supervised machine learning [9] to evaluate network log records and generate results. He accessed the UCI machine learning library [10] and utilized the K-means and self-organizing feature map (SOFM) methods to achieve 97.2% accuracy.

To regulate traffic based on repeated stem analysis of actions on a firewall device, Al-Behadili et al. [11] implemented a machine-learning system. Meanwhile, the key factors output categorization model is inaccurate. For this, he turned to multiclass categorization through decision tree analysis. The author recommended building the categorization model using the divide-and-conquer strategy. At the end, he demonstrated a measurement of output and a comparison of classification error.

With an accuracy of 98.5% and cross-entropy damage of 0.022 attenuations of 3-class classifier, Al-Haija et al. [12] in his implementation of a classification model in the firewall systems produced proper action for every communicated packet by trying to analyze packet attributes using shallow neural network (SNN). The model evaluated by multiple evaluation metrics.

Four different support vector machine algorithms were applied to the dataset by Ertam et al. [13], and the resulting output was a comparison of the algorithms' accuracy using an activation function and a receiver operating characteristic (ROC) curve.

Anomaly identification in enormous log files was a topic that was tackled by Hommes et al. [14]. They use label propagation as a semi-supervised learning strategy because of the limited quantity of labeled data available. They employed a real-world dataset from an Internet service provider as its base, and their main goal was to find unusual entries in the firewall's logs.

Using machine learning and excellent quality computing techniques, Ucar et al. [15] suggested an automated methodology for detecting anomalies in the firewall rule library. Many classification algorithms were employed to examine and extract features from the dataset, and the optimal performance was analyzed.

Through data preprocessing treatment (DPT), Amar et al. [16] organize input log files, predicts missing information, and apply a weighted conversion to make it simpler to distinguish malicious actions. Based on the proven success of deep learning in areas such as high-dimensional data analysis, feature selection, and intrusion detection, we present a weighted long short-term memory (WLSTM) deep learning architecture.

TABLE I. COMPARISON BETWEEN PREVIOUS WORK

| Ref | Contributions | Dataset | Algorithms used | Best Accuracy |
|---|---|---|---|---|
| [17] | Create two-class classifiers that can differentiate between the regular and the analogous | Network data is made available to the public through the UCI ML repository. | K-means and SOFM | 97.2% |
| [18] | To monitor for suspicious activity in a network, it's a good idea to implement a Spark-based data security platform. | Confidential multi-source, multi-heterogeneous network log data | Random Forest, JRip algorithm. | 99.9% |
| [19] | Evaluate the efficacy of three different activation functions for multiclass SVMs. | Firat University network log 65k | SVM and RBF activation functions. | 98% |
| [11] | Predict the next move with multiclass ML. | Firat University network log 65k | DT, SVM, ANN, PSO, and ZeroR | 99% |
| [20] | Analyze network logs with ML classifiers. | Private data 500,000 instances | NB, KNN, and J48. | 99% |

### III. METHODOLOGY

#### A. Data Gathering Module

In this study, all logs obtained with the firewall device on UCI dataset [10]. In the dataset, there are 65532 number of instances with 12 features in total. Attribute" Action" is used as a class with four classes named allow, drop, deny, and reset-both. There are four classes in the class-based action attribute. These are the Allow, Deny, Drop, and Reset-both classes. The visual data description has shown in Fig. 1. On the other hand, Fig. 2 shows the dataset class instances visualization.

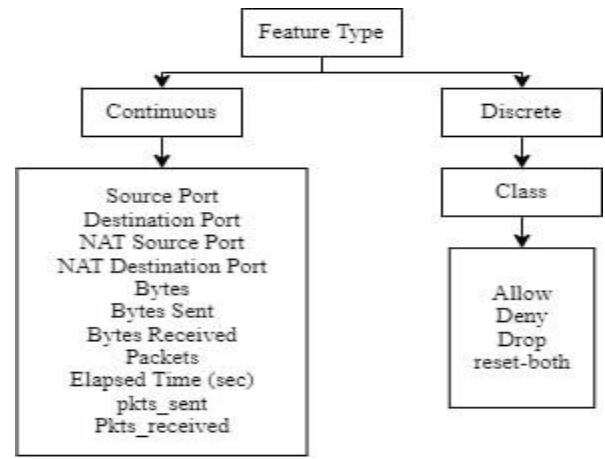

Fig. 1. Data Description with classes

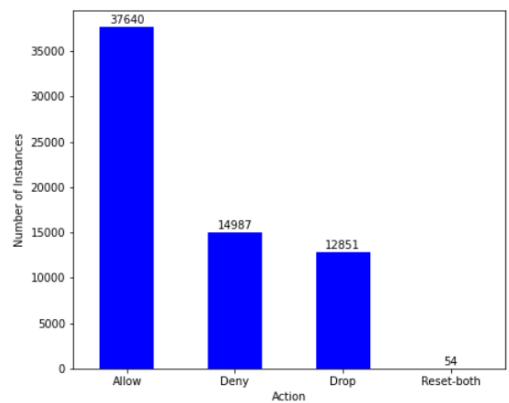

Fig. 2. Total Dataset Class Instances Visualization

*B. Data Pre-Processing*

In this work, we have collected data from UCI open-source platform. We have collected 65,478 firewall logs for this experiment. Classification models can't be used without first cleaning the data and making sure there are no mistakes. Using clean, well-structured data in analyses is dependent on taking the time to reformat and clean datasets. The following steps were taken to process the data we collected:

*1) Data Transformation*

Modifying the dataset is a crucial part of developing a good classification model. Processing information entails preparing it for consumption by doing tasks such as data cleaning, editing, and reduction. The dataset was subjected to a typical scalar transformation, which equated the variance of each feature or variable to one. We have improved performance on this dataset by re-scaling the features independently on the training set and the testing set, without distorting the range of the values, and then using those two sets as one. Afterward, the dataset was converted into a matrix of features including samples (12x65532) and a vector of labels (1x65532).

*2) Data Labeling*

To help a machine learning model understand the context of raw data by recognizing it and labeling it with relevant and useful terms. Due to the categorical nature of the dataset class features, the data must be encoded into numerical labels (labeling) before it can be analyzed mathematically by algorithms for machine learning and calculation. For each target class listed in Table II, we used one hot encoding approach [21] to provide accurate labels.

TABLE II. INDEXED CLASS OF THE DATASET

| *Class* | *Indexed* |
|---|---|
| allow | 0 |
| deny | 1 |
| drop | 2 |
| Reset-both | 3 |

*3) Dataset Randomization*

To reduce the possibility of classification bias and enhance the quality of the validation and testing phases, this step involves redistributing dataset samples in a random sequence. We did this by employing shuffling as a randomization scheme for our datasets, which causes data samples to be distributed arbitrarily over the available spaces.

*4) Dataset Splitting Up*

At this point, we split the data into two parts: the training data and the test data. The full dataset was split in two for the sake of algorithmic application: a smaller, testing dataset of 19,660 cases and a larger, training dataset of 70%, including 45,872 occurrences.

*C. Machine Learning Model*

In this work, we have applied 4 machine learning algorithms to get our result. Support Vector Machine, K- Nearest Neighbor, Random Forest, and Logistic Regression to train and classify the communication traffic records provided by the open-source UCI dataset.

*1) Random Forest (RF)*

Random Forest is a powerful and widely-applied classification and regression algorithm. It is an ensemble approach that combines multiple independent decision trees to produce a model that is more resilient and accurate. Random Forest is based on the idea of constructing a forest of uncorrelated decision trees and average their predictions to get a conclusion [22]. On the basis of the values of the independent variables, a decision tree subdivides the data into increasingly smaller groupings. At each split, the algorithm picks the variable and split point that, depending on the type of decision tree applied, result in the greatest variance or impurity reduction. Each node indicates a data split and each leaf represents a forecast in the final output, which is a tree structure. Random Forest augments the decision tree approach by introducing randomness. During each split, just a random selection of variables, as opposed to all variables, is assessed. In addition, each tree is constructed using a random subset of the data, known as a bootstrapped sample, rather than the entire dataset. These two randomization sources ensure that the trees are uncorrelated and that each tree receives a distinct data sample. The ultimate prediction provided by a Random Forest [23] is the mean of the individual tree forecasts. This averaging has a regularizing effect, reducing the forecast variance and improving the model's consistency. Random Forest also provides a measure of feature significance, allowing you to identify which variables have the most impact on the outcome.

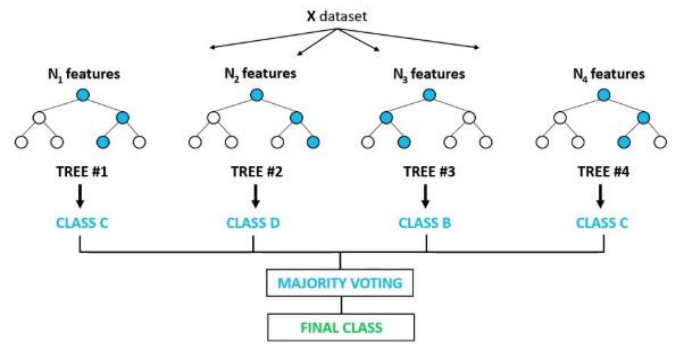

Fig. 3. Visualization of random forest [24]

*2) Logistic Regression (LR)*

Logistic Regression [25] is a helpful statistical method for analyzing datasets in which the outcomes are determined by a variety of factors that are independent of one another. Binary categorization is applied when the dependent variable may only take on one of two values, "yes" or "no." Logical regression employs the logistic function, a subset of the sigmoid family of functions that maps any input to the interval [0,1], to express the relationship between the dependent and independent variables. As it maps any input to the interval [0,1], the output of the logistic function may be interpreted as the probability that the outcome is 1. Often, a threshold value is utilized to determine if the probability is above or below the threshold, enabling a binary forecast. For instance, if the threshold is set to 0.5, an observation is labeled "yes" if the probability is greater than 0.5, and "no" otherwise. See Fig. 4. for an explanation of logistic regression.

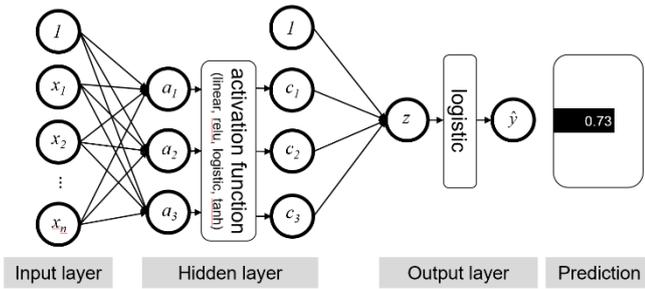

Fig. 4. Basic operation of logistic regression [26]

*3) K- Nearest Neighbor (KNN)*

K-Nearest Neighbor (KNN) [27] is a technique for classification and regression applications that is non-parametric. The algorithm belongs to the class of supervised learning techniques, which means that it learns to make predictions based on labeled data. In KNN, the prediction for a new input data point is derived from the K training data points that are closest to it. K is a hyperparameter whose value must be supplied before training the model. The procedure is known as the Closest Neighbor algorithm when K is set to 1. The result of the KNN method for classification problems is the class label that appears most frequently among the K nearest neighbors. In other words, the method allocates the new input point to the class with the highest representation among its K nearest neighbors. The result for regression tasks is the mean or median of the K nearest neighbors. KNN [28] may also be applied to classification situations involving more than two potential classes. In such situations, the KNN algorithm can employ strategies such as the one-vs-all strategy, in which it trains numerous binary classifiers for each class, and the one-vs-one approach, in which it learns a classifier for each pair of classes. Fig 5 shows the basic organizing concept of KNN.

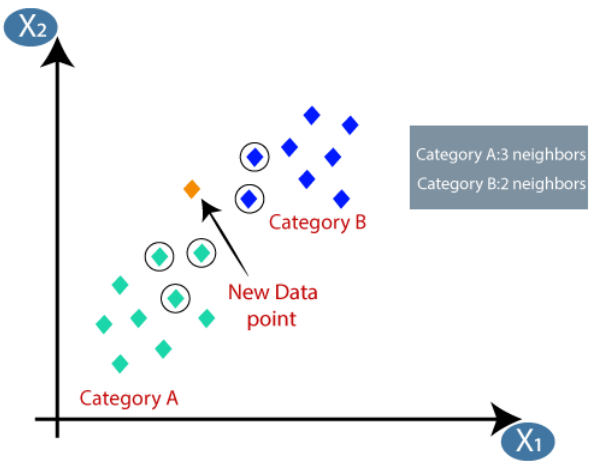

Fig.5. Basic structure of KNN [28]

*4) Support Vector Machine (SVM)*

Support Vector Machines (SVMs) [29] are an effective machine-learning technique used for classification and regression. SVMs are a technique for supervised learning that may be used in both linear and nonlinear situations. An SVM seeks to identify the optimal hyperplane that divides the data into two or more classes. The hyperplane is the decision boundary that maximizes the difference between classes. The margin is defined as the distance between each class's nearest data points and the hyperplane. The SVM method seeks to maximize the margin, which improves the model's ability to generalize to new data. In situations when a linear decision boundary cannot be used to separate the data, SVMs translate the data to a higher dimensional space, where a linear boundary may be used to separate the data, using a method called the kernel trick. This method is known as kernelization [30], and the kernel function employed may be linear, polynomial, radial basis function (RBF), sigmoid, or any other acceptable kernel function. SVMs function by determining the optimal regularization parameter (C) and kernel parameter values (if using a non-linear kernel). The regularization parameter regulates the trade-off between the maximization of the margin and the minimization of misclassification mistakes. A small value of C increases the margin but may increase the number of misclassifications, whereas a big value of C may result in overfitting. SVMs can also be applied to situations involving the classification of many classes. In such situations, the algorithm may employ approaches such as one-vs-one or one-vs-all to train several binary classifiers for each class or a single multiclass classifier.

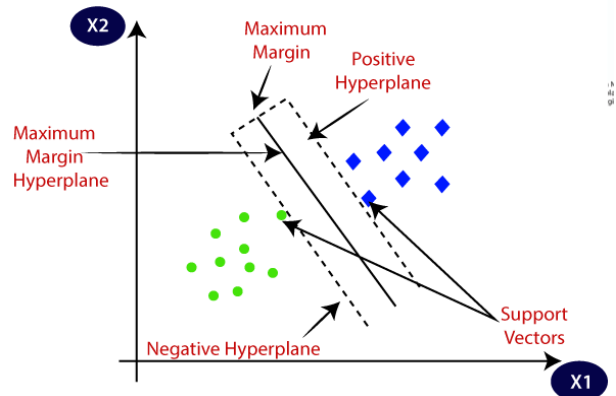

Fig. 6. Basic working procedure of SVM [29]

### IV. RESULT & DISCUSSION

Several models were constructed with Machine learning for use in the studies. Models utilizing ML methods were developed with Weka 3.8.5, while the DL model was developed with Python 3.8 in a Google Collab notebook environment. Experiments were run on 65,478 firewall logs instances, and a target class of "Accept," "Drop," "Deny," and "Reset-Both" was employed. Different characteristics were employed in each trial. The experiment employed the same 11 attributes as the first, plus Application and Category, for a total of 13. Models were trained and constructed with 10-fold cross-validation in both trials.

For reducing the workload on security analyst and providing a reliable and efficient way to automate the classification of network log files, we used these machine learning model. Additionally, we analyzed the important classification metric to avoid misclassification on the log files which can have a serious consequences for network security. As network log file classification is an important ongoing area of research due to evolving of cyber threats, the importance of developing new strategies and techniques of classification is necessary. This research showed all the possible outcomes to develop the area.

*A. Result*

Accuracy, Recall, Precision, $F_1$-Scores, were used to assess the multiclass machine learning models. Table III demonstrates the optimal parameters for both of the trials.

TABLE III.  COMPARATIVE ANALYSIS OF FOUR MACHINE LEARNING TECHNIQUES

| Methods | Class | Precision | Recall | $F_1$-Score | AUC |
|---|---|---|---|---|---|
| SVM | allow | 0.98 | 0.95 | 0.97 | 0.95 |
|  | drop | 0.98 | 0.88 | 0.93 |  |
|  | deny | 0.85 | 1 | 0.92 |  |
|  | reset-both | 1 | 0 | 0 |  |
| RF | allow | 1 | 1 | 1 | 0.99 |
|  | drop | 0.99 | 0.96 | 0.97 |  |
|  | deny | 0.96 | 1 | 0.98 |  |
|  | reset-both | 1 | 0 | 0 |  |
| LR | allow | 1 | 0.99 | 0.99 | 0.98 |
|  | drop | 0.98 | 0.95 | 0.97 |  |
|  | deny | 0.93 | 1 | 0.96 |  |
|  | reset-both | 1 | 0 | 0 |  |
| KNN | allow | 1 | 1 | 1 | 0.99 |
|  | drop | 0.99 | 0.99 | 0.99 |  |
|  | deny | 0.99 | 1 | 0.99 |  |
|  | reset-both | 1 | 0 | 0 |  |

The given results show that Random Forest (RF) and K-Nearest Neighbor (KNN) algorithms have the highest accuracy (99%), followed by SVM (95%) and Logistic Regression (98%). However, accuracy alone is not always the best metric for evaluating models. It is important to consider other metrics such as f1 score, precision, and recall, which provide a more comprehensive picture of model performance. Comparing the f1 score, we can see that Random Forest has the highest average f1 score (0.99), followed by KNN (0.98), SVM (0.95), and Logistic Regression (0.97). This indicates that Random Forest has the best balance between precision and recall. In terms of precision, Random Forest and KNN have the highest average precision (0.98), followed by SVM (0.94) and Logistic Regression (0.93). Precision is the ratio of true positives to the total predicted positives, so a high precision score indicates that the model has a low false positive rate. In terms of recall, Random Forest and RF have the highest average recall (0.99), followed by SVM (0.95) and Logistic Regression (0.95). Recall is the ratio of true positives to the total actual positives, so a high recall score indicates that the model has a low false negative rate. Overall, the results suggest that Random Forest and KNN are the most effective algorithms in terms of accuracy, $F_1$ − score, precision, and recall. SVM and Logistic Regression also performed well but had slightly lower scores in some of these metrics. The choice of the best algorithm will depend on the specific task at hand and the trade-offs between different performance metrics.

Let's check out the output of these four algorithms, and we can make an informed judgment from here.

If we analyze the confusion matrices [31], we can see that the "allow", "deny" and "drop" are classified correctly with high precision and recall score. Due to the dataset boundary and lower amount of data in "reset both" is the reason of lower accuracy in that specific class. In all the algorithm "reset both" had lower recall score which identify that this class was the most difficult one to classify correctly for all algorithm. Overall, random forest is the most effective algorithm for classifying for the log file analysis and there is a room for improvement in the classification for "reset both" class.

*1) Support Vector Machine*

```
              precision    recall  f1-score   support

       allow       0.98      0.95      0.97      3470
        drop       0.98      0.88      0.93      1297
        deny       0.85      1.00      0.92      1229
  reset-both       1.00      0.00      0.00         4

    accuracy                           0.95      6000
   macro avg       0.95      0.71      0.70      6000
weighted avg       0.95      0.95      0.95      6000
```

Fig.7. Classification Report of SVM

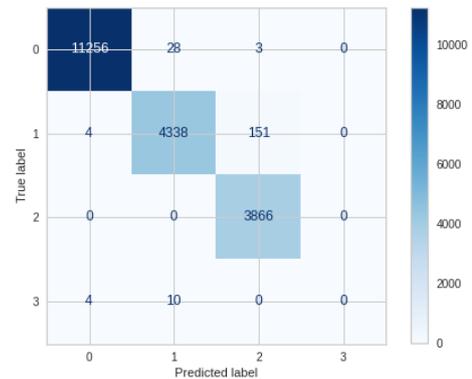

Fig. 8. Confusion Matrix of SVM

*2) Logistic Regression*

```
              precision    recall  f1-score   support

       allow       1.00      0.99      0.99      3470
        drop       0.98      0.95      0.97      1297
        deny       0.93      1.00      0.96      1229
  reset-both       1.00      0.00      0.00         4

    accuracy                           0.98      6000
   macro avg       0.98      0.73      0.73      6000
weighted avg       0.98      0.98      0.98      6000
```

Fig.9. Classification Report of LR

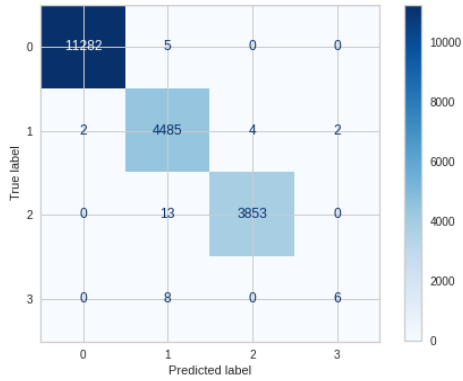

Fig. 10. Confusion Matrix of LR

*3) K- Nearest Neighbor*

```
              precision    recall  f1-score   support

       allow       1.00      1.00      1.00      3470
        drop       0.99      0.99      0.99      1297
        deny       0.99      1.00      0.99      1229
  reset-both       1.00      0.00      0.00         4

    accuracy                           0.99      6000
   macro avg       0.99      0.75      0.74      6000
weighted avg       0.99      0.99      0.99      6000
```

Fig. 11. Classification Report of KNN

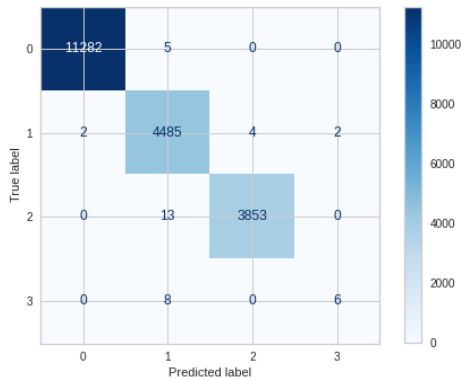

Fig. 12. Confusion Matrix of KNN

*4) Random Forest*

```
              precision    recall  f1-score   support

       allow       1.00      1.00      1.00      3470
        drop       0.99      0.96      0.97      1297
        deny       0.96      1.00      0.98      1229
  reset-both       1.00      0.00      0.00         4

    accuracy                           0.99      6000
   macro avg       0.99      0.74      0.74      6000
weighted avg       0.99      0.99      0.99      6000
```

Fig.13. Classification Report of RF

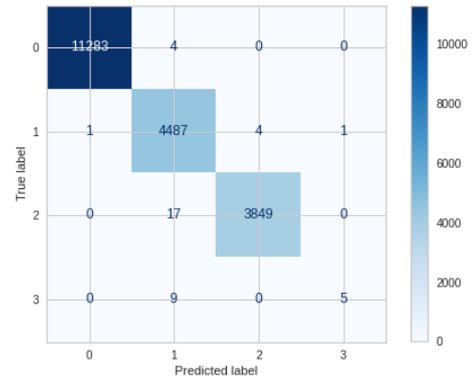

Fig. 14. Confusion Matrix of RF

## V. CONCLUSIONS AND FUTURE WORK

Being the initial line of defense, firewalls play a crucial role in protecting a company's network. Firewalls offer protection not just from outsiders but also from malicious insiders. Considering the significance of firewalls to system security, this research aimed to develop many Machine Learning models capable of categorizing sessions in firewall logs and determining the appropriate response. To determine if the action was "Accept," "Drop," "Deny," or "Reset-both," a comparison of five multiclass methods was conducted. We utilized a publicly available dataset including 65,478 firewall logs for training and assessing our models. Also, the study demonstrated a comparison of the characteristics by experimenting with a single set of features. All told, 11 characteristics were tested for in this study. The tests demonstrated the effectiveness of factoring in a website's application and category when deciding on what action to take, thereby enhancing the firewall's efficiency. To top it all off, the proposed models in this study all achieved great accuracy, with the RF algorithm achieving a maximum of 99% accuracy in the experiment. Our research lends credence to the notion that ML methods may be used to reliably and swiftly automate the classification of firewall data, hence enhancing the security of enterprise networks. In addition, the findings can aid in the development of new strategies to counteract cyber-attacks and enhance the security and protection offered by firewalls and antivirus software. In the future, we will work on this experiment with a more organized and bulk dataset.